\documentclass[aps,prb,floatfix,showpacs,twocolumn,groupedaddress]{revtex4}

\usepackage{color}
\usepackage{bm}
\usepackage{amssymb}
\usepackage{amsmath}
\usepackage{amstext}
\usepackage{graphics}
\usepackage{epsfig}
\usepackage{placeins}
\usepackage{psfrag}
\usepackage{subfigure}

\begin{document}

\preprint{ Version \today}

\title{Current-Driven Domain-Wall Dynamics in Curved Ferromagnetic Nanowires}

\author{Benjamin Kr\"uger}
\author{Daniela Pfannkuche}

\affiliation{I. Institut f\"ur Theoretische Physik, Universit\"at Hamburg,
  Jungiusstr. 9, 20355 Hamburg, Germany}

\author{Markus Bolte}
\author{Guido Meier}
\author{Ulrich Merkt}

\affiliation{Institut f\"ur Angewandte Physik und Zentrum f\"ur Mikrostrukturforschung, Universit\"at Hamburg,
  Jungiusstr. 11, 20355 Hamburg, Germany}

\date{\today}

\begin{abstract}
The current-induced motion of a domain wall in a semicircle nanowire with
applied Zeeman field is investigated. Starting from a micromagnetic model we
derive an analytical solution which characterizes the domain-wall
motion as a harmonic oscillation. This solution relates the micromagnetic
material parameters with the dynamical characteristics of a harmonic
oscillator, i.e., domain-wall
mass, resonance frequency, damping constant, and force acting on the wall. For wires with strong curvature the dipole moment of the wall as
well as its geometry influence the eigenmodes of the oscillator.
Based on these results we suggest experiments for the determination of
material parameters which otherwise are difficult to access. Numerical
calculations confirm our analytical solution and show its limitations.
\end{abstract}

\date{\today}

\pacs{75.60.Ch, 72.25.Ba, 76.50.+g}

\maketitle

\section{Introduction}
\label{Introduction}
Field-driven dynamics of magnetic domain walls have been intensely studied over the
last decades. \cite{Thiele1974,Walker1974} The topic has recently
regained interest by the discovery that spin-polarized currents of high density
can alter magnetization configurations \cite{Slonczewski1996,Berger1996,Myers1999,GrollierAPL2003} and move domain walls.
\cite{Yamaguchi2004,Koo2002,Saitoh04,KlaeuiAPL2003,VernierEPL2004} Current-induced magnetic switching is viewed as
a promising solution for the realization of magnetic random access memories
\cite{KatinePRL2000,GrollierAPL2003,KrivorotovScience2005}, while
current-induced domain-wall motion has potential application in spintronic data
storage devices, e.g. in the racetrack memory \cite{ParkinPatent2004} or data transfer
schemes. \cite{AllwoodScience2002,AllwoodScience2005,CowburnPatent2004} Several
models of current-driven magnetization dynamics have been established to
explain the electronic origin of current-induced magnetization changes and to predict their effects.
\cite{Slonczewski1996,Berger1996,TataraPRL2004,ThiavilleEPL2005,WaintalEPL2004,ZhangLiPRL2004}
At first it was assumed that for
finite domain walls the spins of the conduction electrons adiabatically follow the
local magnetic moments. \cite{TataraPRL2004,ZhangLi2004a} Later the theoretical
model was extended to include a non-adiabatic mismatch between the
current polarization and the direction of magnetization.
\cite{WaintalEPL2004,ThiavilleEPL2005,ZhangLiPRL2004}

The measured and calculated velocities of current-driven magnetic domain walls
in thin nanowires vary by several orders of magnitude even for the same material.
\cite{KlaeuiPRL2005,Yamaguchi2004,ZhangLiPRL2004,OnoScience1999,ThiavilleNatureMat}
While it has been originally suggested that the discrepancy could be due to thermal activation
\cite{VernierEPL2004,ThiavilleNatureMat,KlaeuiPRL2005a,YamaguchiAPL2005,
YamaguchiErratum2006} or surface roughness \cite{ThiavilleNatureMat},
it has recently been found that the domain-wall velocity depends on the type of the domain wall
\cite{KlaeuiPRL2005a,HePRB2006} which can be changed by a spin-polarized
current. \cite{Koo2002,KlaeuiPRL2005a,KlaeuiAPL2006,ourALSdata} Recently it
has been observed that the
velocity of field-driven domain-wall motion \cite{AtkinsonNature2003} can
be altered by $\pm$ 100 m/s
by a pulsed spin-polarized current \cite{Hayashi2006} and that the motion
can even be halted completely. \cite{Thomas2006} It is now assumed
that the adiabatic term is largely responsible for the acceleration of the
domain wall while the non-adiabatic term will cause the wall to continually move.
\cite{ZhangLiPRL2004} It has been shown that domain-wall oscillations excited with an ac current 
at its resonance frequency require current densities one to two orders of magnitude less ($10^{10}~\mbox{A}/\mbox{m}^{2}$, see Ref. 
\cite{Saitoh04,TataraAPL2005}) than for pulsed excitations ($10^{11}-10^{12}~ \mbox{A}/\mbox{m}^{2}$, see Ref.
\cite{KlaeuiPRL2005a,Yamaguchi2004,GrollierAPL2003,VernierEPL2004}).

Here we show that the harmonic-oscillator model follows naturally from a
micromagnetic model that describes the excitation of transverse walls in thin narrow rings.
\cite{McMichaelIEEE,KlaeuiPhaseDiagram,ThiavilleJMMM}
Solving analytically  the Landau-Lifshitz-Gilbert  
equation extended by the current corrections due to Zhang and Li
\cite{ZhangLiPRL2004} we are able to express the properties of the driven
oscillator by the quantities determining the micromagnetic model. A comparison
of the numerical calculations with our analytical solution confirms the importance of the geometry
due to the curved wires. 
Finally, we suggest experiments which can determine the values of the non-adiabatic
spin torque and the Gilbert damping parameter.

\section{Model}
\label{Model}
\begin{figure}
  \includegraphics[angle = 270, width = 1.0\columnwidth]{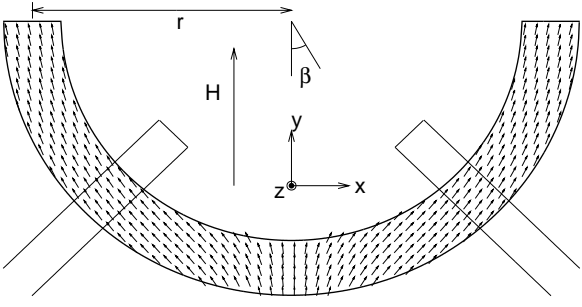}
  \caption{Scheme of the semicircle nanowire with radius $r$ in a magnetic field $H$. The static magnetization in the absence of a current is indicated by small arrows. The two rectangles under angles $\beta = \pm 45^{\circ}$ are the electrical contacts. \label{model}}
\end{figure}
Figure \ref{model} shows a ferromagnetic semicircle
nanowire with a domain wall at its bottom placed in an
external magnetic field. \cite{Preparation} The wall is excited by an
oscillating
current flowing between the two contacts. \cite{Saitoh04}

The magnetization dynamics of a magnetic wire is well described by the
Landau-Lifshitz-Gilbert (LLG) equation. \cite{LLG} In the presence of a
spin-polarized current density $\vec{j}$, the interaction between the itinerant
electrons and 
the magnetization $\vec{M}$ leads to an extension of the LLG equation.
This extension was derived from a quantum mechanical model by Zhang and Li.
\cite{ZhangLiPRL2004} Their semiclassical approximation results in
the extended LLG equation (in Gilbert form)
\begin{equation}
\begin{split}
\frac{d \vec M}{d t} =
 &- \gamma \vec M \times \vec  H_{\mbox{eff}} + \frac{\alpha}{M_s} \vec M \times
\frac{d \vec M}{d t}\\
 &- \frac{b_j}{M_s^2} \vec M \times \left( \vec M \times (\vec j \cdot \vec
\nabla) \vec M \right)\\
 &- \xi \frac{b_j}{M_s} \vec M \times (\vec j \cdot \vec \nabla) \vec M
\end{split}
\label{LLG-Zang-Li}
\end{equation}
with the gyromagnetic ratio $\gamma$, the Gilbert damping parameter $\alpha$, the
saturation magnetization $M_s$, and the ratio between
exchange relaxation time and spin-flip relaxation time $\xi =
\tau_{\mbox{ex}}/\tau_{\mbox{sf}}$. The
effective magnetic field $H_{\mbox{eff}}$ includes the external as well
as the internal fields.  In this model the spin current is sensitive to the
spatial inhomogeneities of the magnetization with a coupling constant $b_j =
\frac{P \mu_B}{e M_s (1+ \xi^2)}$ where $P$ denotes the spin polarization of
the current and $\mu_B$ is the Bohr magneton.

Since the saturation magnetization is constant for a given material at fixed
temperature, $\vec M$ is perpendicular to $\frac{d \vec M}{d t}$ and Eq.
(\ref{LLG-Zang-Li}) can be reformulated to an explicit equation of motion for
the magnetization 
\begin{equation}
\begin{split}
\frac{d \vec M}{d t} =
 &- \gamma' \vec M \times \vec  H_{\mbox{eff}} -
\frac{\alpha \gamma'}{M_s} \vec M \times \left( \vec M \times \vec 
H_{\mbox{eff}} \right)\\
 &- \frac{b_j'}{M_s^2} (1+\alpha \xi) \vec M \times \left(
\vec M \times (\vec j \cdot \vec \nabla) \vec M \right)\\
 &- \frac{b_j'}{M_s} (\xi - \alpha) \vec M \times (\vec j
\cdot \vec \nabla) \vec M
\end{split}
\label{LLG-Zang-Li2}
\end{equation}
with the abbreviations $\gamma' = \frac{\gamma}{1+\alpha^2}$ and $b_j' = \frac{b_j}{1+\alpha^2}$.
This equation is the starting point for the analytical as well as for the
numerical calculations presented in the following.

\section{Analytical Calculations of the Straight Wire}
\label{Analytic Calculations}

For the analytical treatment of Eq.~(\ref{LLG-Zang-Li2}) we transform the semicircle wire in a homogeneous Zeeman field to a straight wire in a spatially variing field. The wire is directed along the $x$-axis. The direction of the magnetization will be expressed in a polar
spin basis $\vec{M} = M_s
(\cos \theta ,\sin \theta \cos \phi ,\sin \theta \sin \phi)$. 
In the absence of electric current and external magnetic field the energy of a
domain wall within the wire is
\begin{equation}
E = S \int \left[ A \left( \frac{\partial \theta(x)}{\partial x} \right)^2 + K
\sin^2 \theta(x) \right] dx,
\label{E_neel}
\end{equation}
where $\theta$ denotes the angle between the wire axis and the magnetization.
$A$ and $K$ denote the exchange and the shape
anisotropy constant. This functional can be minimized by the well known N\'{e}el wall described by the angle
\begin{equation}
\theta = \pi - 2 \arctan{\left( e^{\frac{x-X}{\lambda}} \right)}.
\label{Neel}
\end{equation}
The center of the wall is at position $X$ and the width of the domain wall is 
$\lambda= \sqrt{A/K}$. 
From Eq. (\ref{Neel}) two expressions
\begin{equation}
\cos \theta  = \tanh \left(\frac{x-X}{\lambda} \right)
\; \;, \;\;
\sin \theta  = \frac{1}{\cosh \left(\frac{x-X}{\lambda} \right)}
\label{neel_sincos_theta}
\end{equation}
can be derived which will be useful in our further calculations.

In the presence of an external field $H_{\mbox{ext}}$ the demagnetization energy
$K_{\perp} \sin^2 \theta \sin^2 \phi$  caused by the rotation of the wall around the wire axis can no longer be neglected. We include the external field perpendicular to the wire into the shape anisotropy $K_{\perp}$.
The energy functional in Eq. (\ref{E_neel}) has to be extended to
\begin{equation}
\begin{split}
E & = \int \left[ K \sin^2 \theta + A \left( \frac{\partial \theta}{\partial
x} \right)^2 \right] dV\\
 & + \int \left[ K_{\perp} \sin^2 \theta \sin^2 \phi - \mu_0 M_s
H_{\mbox{ext}}(x) \cos \theta  \right] dV.
\end{split}
\label{neel_energy}
\end{equation}
Here we have restricted ourselves to an external field parallel to the wire.
Also the crystalline anisotropy has been
neglected. \cite{Anisotropy} From the energy functional in Eq. (\ref{neel_energy})
we derive the effective magnetic field through the relation 
$\vec{H}_{\mbox{eff}}=-\frac{1}{\mu_0}\frac{\delta E}{\delta \vec{M}}$. 

From the extended LLG equation (\ref{LLG-Zang-Li2}) in the polar spin basis we obtain
\begin{equation}
\begin{split}
\dot \theta & = -\frac{\gamma'}{\mu_0 M_s \sin(\theta)} \frac{\delta E}{\delta
\phi} - \frac{\gamma' \alpha}{\mu_0 M_s} \frac{\delta E}{\delta \theta}\\
 & + b_j' (1+ \alpha \xi) \vec j \cdot \vec \nabla \theta + b_j' (\xi - \alpha)
\sin(\theta) \vec j \cdot \vec \nabla \phi
\end{split}
\label{angular_LLG_theta}
\end{equation}
and
\begin{equation}
\begin{split}
\dot \phi \sin \theta & = \frac{\gamma'}{\mu_0 M_s} \frac{\delta E}{\delta
\theta} - \frac{\gamma' \alpha}{\mu_0 M_s \sin(\theta)} \frac{\delta E}{\delta
\phi}\\
 & + \sin(\theta) b_j' (1+ \alpha \xi) \vec j \cdot \vec \nabla \phi - b_j' (\xi - \alpha)
 \vec j \cdot \vec \nabla \theta.\\
\end{split}
\label{angular_LLG_phi}
\end{equation}

Assuming that the moving wall stays a N\'{e}el wall we can describe its motion,
following the description of Schreyer and Walker \cite{Walker1974}, by two
dynamical variables: the position of its center $X$
and its angle around the wire axis $\phi(x) = \phi$ that is uniform along the 
wire. 
With Eq.~(\ref{neel_sincos_theta}) and Eq.~(\ref{neel_energy}) we get from
Eq.~(\ref{angular_LLG_theta}) and Eq.~(\ref{angular_LLG_phi})
\begin{equation}
\begin{split}
\frac{\sin(\theta)}{\lambda} \dot X & = -\frac{2 K_{\perp} \gamma' }{\mu_0 M_s}
\sin(\theta) \sin(\phi) \cos(\phi)\\
 & - \alpha \gamma' \sin(\theta) H_{\mbox{ext}}(x) - \frac{b_j'}{\lambda}
\sin(\theta) (1+\alpha \xi) j\\
 & - \frac{2 K_{\perp} \gamma' \alpha}{\mu_0 M_s} \sin(\theta) \cos(\theta)
\sin^2(\phi)\\
\end{split}
\label{x_dot}
\end{equation}
and
\begin{equation}
\begin{split}
\dot \phi \sin \theta & = \sin(\theta) \gamma' H_{\mbox{ext}}(x)+ \sin(\theta)
\frac{b_j' (\xi - \alpha) j}{\lambda}\\
 & -2 \sin(\theta) \gamma' \alpha K_{\perp} \sin(\phi) \cos(\phi) \frac{1}{\mu_0
M_s}\\
 & + \frac{2 K_{\perp} \gamma'}{\mu_0 M_s} \sin(\theta) \cos(\theta)
\sin^2(\phi).\\
\end{split}
\end{equation}
Note, that $X$ and $\phi$ depend on the position $x$ along the
wire. In the following we show that a
solution consistent with our initial assumptions exists for small exitations.
\cite{Walker1974} Note, that this condition holds for realistic current
densities.

Assuming that $H_{\mbox{ext}}(x)$ varies slowly on the length scale of the domain-wall width $\lambda = \sqrt{A/K}$, $\sin \theta $ is replaced by a $\delta$-function $\pi
\lambda \delta(x-X)$ in view of Eq. (\ref{neel_sincos_theta}). Also we neglect terms which are nonlinear in $\phi$. This approximation
holds for angles $\phi$ smaller than about $10^{\circ}$.

The equations of motion for the domain wall then become 
\begin{equation}
\dot X = - \lambda 2 \gamma' K_{\perp} \phi \frac{1}{\mu_0 M_s} - \lambda
\gamma' \alpha H_{\mbox{ext}}(X)- b_j' (1+\alpha \xi)j
\label{X_dot_straight_wire}
\end{equation}
and
\begin{equation}
\dot \phi = \gamma' H_{\mbox{ext}}(X) -2 \gamma' \alpha K_{\perp} \phi
\frac{1}{\mu_0 M_s}+ \frac{b_j' (\xi - \alpha) j}{\lambda}.
\label{phi_dot_straight_wire}
\end{equation}
These equations are general equations of motion with a time
dependent current density $j$.\\
In the limit of a steady current and a homogeneous magnetic field one can
calculate the initial velocity of the wall by setting $\phi = 0$, the initial
condition of the N\'{e}el wall. This leads
to the initial velocity
$\dot X_i = - \lambda \gamma' \alpha H_{\mbox{ext}}- b_j' (1+\alpha \xi)j$
which is exactly the value obtained by Zhang and
Li. \cite{ZhangLiPRL2004} The terminal velocity $\dot X_f = - \left( \lambda \gamma H_{\mbox{ext}} + b_j \xi j
\right)/\alpha$ is calculated by setting $\dot \phi = 0$, i.e., stationary motion. This velocity is also identical to the one calculated by Zhang and Li. Similar relations have recently been found by Dugaev et al. \cite{Dugaev06}

The domain-wall mass is obtained by determining $\sin \phi$ in the absence of
electric currents and external fields. From Eq. (\ref{X_dot_straight_wire})
we obtain for the stationary motion
\begin{equation}
\phi = - \dot X \frac{\mu_0 M_s}{2 \lambda \gamma'
K_{\perp}}.
\end{equation}
Inserting this result into the $\phi$-dependent part of the domain-wall energy in Eq. (\ref{neel_energy}) and comparing with the energy $E$ of 
the domain-wall quasiparticle
\begin{equation}
\begin{split}
\frac{1}{2} m \dot X^2 & = E\\
 & = S \int dx \; K_{\perp} \sin^2(\theta) \left( \dot X \frac{\mu_0
M_s}{\lambda 2 \gamma' K_{\perp}} \right)^2\\
 & = \frac{1}{2} \frac{S \mu_0^2 M_s^2}{\lambda \gamma'^2 K_{\perp}} \dot X^2\\
\end{split}
\end{equation}
we arrive at the domain-wall mass
\begin{equation}
m = \frac{S \mu_0^2 M_s^2}{\lambda \gamma'^2 K_{\perp}}.
\label{wall_mass}
\end{equation}
Note that this result relates the phenomenological domain-wall mass of a
tail-to-tail N\'{e}el wall to the micromagnetic material parameters. 

In the case of a curved wire the projection of a uniform external field along
the wire is given by $H_{\mbox{ext}}(x) = H_0 \sin \left(x/r \right)$. Transferring this to our straight wire model, at small displacements of 
the domain wall ($X << r$) the wall is exposed 
to the external field  $H_{\mbox{ext}} \approx H_0 X/r$. Then the
equations of motion become a system of two coupled linear differential equations
of first order:
\begin{equation}
\begin{split}
\left( \begin{matrix}
\dot X\\
\dot \phi\\
\end{matrix} \right)
& = \gamma'
\left( \begin{matrix}
-\lambda \alpha H_0 \frac{1}{r} & -\lambda 2 K_{\perp}
\frac{1}{\mu_0 M_s}\\
 H_0 \frac{1}{r} & -2 \alpha K_{\perp} \frac{1}{\mu_0 M_s}\\
\end{matrix} \right)
\left( \begin{matrix}
X\\
\phi\\
\end{matrix} \right)\\
 & + b_j' j
\left( \begin{matrix}
- (1+\alpha \xi)\\
\frac{(\xi - \alpha) }{\lambda}\\
\end{matrix} \right).\\
\end{split}
\label{ode_system_curved_wire}
\end{equation}
Except for the non-vanishing first matrix element $-\lambda \alpha H_0/r$ these equations are equivalent to those of a driven harmonic oscillator.
For a time dependent current density of the form $j_0 e^{i \Omega t}$ the general
solution
\begin{equation}
\begin{split}
\left( \begin{matrix}
X(t)\\
\phi(t)\\
\end{matrix} \right)
= &
\left( \begin{matrix}
X_+\\
\phi_+\\
\end{matrix} \right)
e^{-\Gamma t + i \omega_f t} +
\left( \begin{matrix}
X_-\\
\phi_-\\
\end{matrix} \right)
e^{-\Gamma t - i \omega_f t}\\
 & +\frac{1}{\omega_r^2 - \Omega^2 + 2 i \Omega \Gamma}
\frac{F}{m}
\end{split}
\label{sol_X(t)_phi(t)}
\end{equation}
consists of an exponentially damped starting configuration with the initial conditions described by $X_{\pm}$ and $\phi_{\pm}$ and a
current-driven oscillation with the driving force $\vec F$. The damping constant
\begin{equation}
\Gamma = \alpha \gamma' \left( \frac{\lambda H_0 }{2r} + \frac{K_{\perp} }{\mu_0
M_s} \right)
\label{oscillator_damping}
\end{equation}
depends on the ratio of applied magnetic field and ring radius. It
represents the restoring force acting on the domain wall. This dependence
of the damping constant on the restoring force expresses that the damping is spatially dependent.
This also leads to a second term in the frequency of the free oscillation
\begin{equation}
\omega_f = \sqrt{\frac{2 \gamma'^2 H_0 \lambda K_{\perp} }{\mu_0 M_s r}-\alpha^2
\gamma'^2 \left( \frac{K_{\perp} }{\mu_0 M_s} - \frac{\lambda H_0 }{2r}
\right)^2}
\end{equation}
which is different from $\Gamma$.
Hence the resonance frequency 
\begin{equation}
\omega_r = \sqrt{ \omega_f^2 + \Gamma^2 } = \sqrt{\frac{2 \gamma'^2 H_0 \lambda K_{\perp}
}{\mu_0 M_s r} (1+ \alpha^2)}
\label{resonance_frequency}
\end{equation}
depends explicitly on the Gilbert damping $\alpha$ and differs from the
resonance frequency of a normal harmonic oscillator
\begin{equation}
\omega_0 = \sqrt{\frac{D}{m}} = \sqrt{\frac{2 \gamma'^2 H_0 \lambda K_{\perp}
}{\mu_0 M_s r}}
\end{equation}
by the factor $\sqrt{1 + \alpha^2}$.
The constant $D$ is given by $D = F_H/X$ where $F_H$ is the force on
the domain wall due to the external magnetic field.
The force
\begin{equation}
\begin{split}
\vec F = &
- m b_j j_0 e^{i \Omega t}
\left(
\frac{2 \gamma' K_{\perp} \xi}{\mu_0 M_s} + \frac{1+\alpha \xi}{1 + \alpha^2} i
\Omega
\right) \vec e_X\\
 & - m b_j j_0 e^{i \Omega t}
\left(
\gamma' H_0 \frac{1}{r} - \frac{\xi -\alpha}{1+ \alpha^2} \frac{i
\Omega}{\lambda}
\right) \vec e_{\phi}\\
\label{force due to current}
\end{split}
\end{equation}
induced by the current depends on the frequency $\Omega$ of the applied
current.

\begin{figure}
  \includegraphics[angle = 270, width = 1.0\columnwidth]{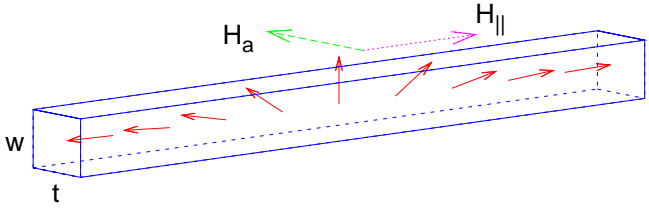}
  \caption{(Color online) Schematic illustration of the magnetization in the N\'{e}el wall (solid
    red arrows) in a straight wire of width $w$ and thickness $t$. $H_{\parallel}$ and $H_a$ are the parallel component of the external field and the anisotropy field, respectively. \label{wall_field}} 
\end{figure}

The terms in Eq. (\ref{force due to current}) can be understood as direct
forces due to the spin torque and the precessions of the magnetization in the external and anisotropy field depicted in Fig.~\ref{wall_field}. The terms proportional
to $i \Omega$ express the current-induced spin torque. They are
the time derivatives of the inhomogenities in Eq.
(\ref{ode_system_curved_wire}). The $H_0$-dependent term is a result of the precession of the magnetization in the
external field which causes a rotation of the wall around the
wire axis. The precession in the anisotropy field, described 
by the $K_{\perp}$-term in Eq. (\ref{force due to current}), causes a
change of the wall velocity.

Except that the force depends on the frequency $\Omega$ of the applied current the result for the domain wall displacement is equal to the one in a harmonic oscillator. With increasing $\Omega$ the force
increases and its phase shifts up to $90^\circ$. In the absence of a
non-adiabatic spin torque ($\xi = 0$) current and 
domain-wall displacement at resonance have opposite sign.
In case of a non-adiabatic torque the phase at resonance frequency between the
current and the magnetization in z-direction is $90^\circ$ when the ratio $\xi$ of exchange and spin-flip relaxation time equals the Gilbert damping parameter ($\xi = \alpha$).
The phase can be used to find out whether a non-adiabatic spin torque
exists and to determine the value of $\xi$ in
comparison to the damping parameter $\alpha$.

The influence of the adiabatic torque on the position of the wall is obtained
by setting $\xi = 0$ in Eq. (\ref{force due to current}). The
$x$-component of the force $\vec F$ due to the adiabatic torque is proportional to
the time 
derivative of the current density. 
Therefore, the adiabatic torque does not accelerate the wall when the current
does not change in time. This explains the observation of Zhang and
Li \cite{ZhangLiPRL2004} that without a non-adiabatic spin torque a domain wall
subjected to a steady current stops moving.
In contrast the non-adiabatic contributions
to the force are proportional to the current density
as well as to its derivative. 

In Eq. (\ref{sol_X(t)_phi(t)}) the starting configuration depends on
$\phi_{\pm}$ and $X_{\pm}$. The equation that follows from
decoupling of Eq. (\ref{ode_system_curved_wire})
\begin{equation}
\phi_{\pm} = \left( \frac{\alpha}{2 \lambda} - \frac{\alpha H_0 \mu_0 M_s}{4r
K_{\perp}} \pm i \frac{\omega_f \mu_0 M_s}{2 \lambda \gamma' K_{\perp}} \right)
X_{\pm}
\end{equation}
connects $\phi_{\pm}$ with $X_{\pm}$. Hence we have two parameters left for
our starting configuration as expected for an oscillation.

With the above analytical model we are able to derive the hitherto
phenomenological oscillator model \cite{Saitoh04} and to express its characteristics by the
micromagnetic material parameters.
Likewise, the measurement of the domain-wall
motion allows the determination of micromagnetic quantities.

\section {Curved Wires}

For curved wires in a homogeneous magnetic field its
component perpendicular to the wire has to be taken into account. Also
the change of the magnetization due to the curvature becomes important.
To include the perpendicular field we calculate the force on the
domain wall as the spatial derivative of its Zeeman energy.
The total magnetic
moments parallel to the wire
\begin{equation}
m_{\parallel} = \int M_s \cos[\theta(x)] dV = - 2 M_s S X
\end{equation}
and perpendicular to the wire
\begin{equation}
m_{\perp} = \int M_s \sin[\theta(x)] dV = \pi M_s S \lambda
\end{equation}
are volume integrals over its magnetization that are readily calculated using the relations in Eq.~(\ref{neel_sincos_theta}). Note that 
$m_{\parallel}$ is the magnetic moment of an abrupt domain wall. With the
magnetic field $H_0$ in $y$-direction the Zeeman energy can be written as
\begin{equation}
E_{\parallel} = \mu_0 M_s H t
\overset{r+\frac{w}{2}}{\underset{r-\frac{w}{2}}{\int}} \; r' \left[
\overset{\beta_0}{\underset{-\frac{\pi}{2}}{\int}} \sin(\beta) d\beta -
\overset{\frac{\pi}{2}}{\underset{\beta_0}{\int}} \sin(\beta) d\beta \right] dr'
\end{equation}
where $\beta_0 = \frac{X}{r}$ is the angle of the position of the domain wall (see Fig. \ref{model}) and $r$, $w$, and $t$ are the radius, the width, and the thickness of the wire. We get
\begin{equation}
E_{\parallel} = - 2 \mu_0 M_s S r H \cos(\beta_0) = 2 \mu_0 M_s S H Y
\label{E_parallel}
\end{equation}
with the cross section $S=wt$. One recognizes that the energy is equivalent to the
energy of a monopole with magnetic charge $Q_M = 2 \mu_0 M_s S$. For small
domain-wall displacements we can write the cosine in Eq.~(\ref{E_parallel}) as a Taylor series up to second order in $X$
and get
\begin{equation}
E_{\parallel} \approx - 2 \mu_0 M_s S r H \left( 1-\frac{X^2}{2r^2} \right).
\end{equation}
The monopole has been included in the above calculations as well as in the calculations
of Saitoh et al. \cite{Saitoh04}
The perpendicular magnetization contributes to the Zeeman energy 
\begin{equation}
E_{\perp} = -\mu_0 m_{\perp} H \cos \left( \frac{X}{r} \right) \approx - P_M H \left( 1-\frac{X^2}{2r^2} \right)
\end{equation}
and can be interpreted as the energy of a magnetic dipole with moment $P_M = \mu_0
\pi M_s S \lambda$.
The Zeeman energy of the perpendicular magnetization has previously not been included
in the magnetic energy. It gives a correction to the magnetic force on the
domain wall, 
\begin{equation}
\begin{split}
F_x & = - \frac{dE}{dX} \approx - \frac{2 \mu_0 M_s S H}{r}X - \frac{\pi \mu_0 M_s S \lambda H}{r^2}X\\
 & = - \frac{Q_M H}{r}X - \frac{P_M H}{r^2}X = - \frac{Q_M}{r}X H \left( 1 +
\frac{\pi \lambda}{2 r} \right).\\
\end{split}
\end{equation}
Thus, we include the action of a field component perpendicular to the wire by
replacing the field in Eq. (\ref{sol_X(t)_phi(t)}) by an effective
field $H_{\mbox{eff}} = H \left( 1 + P_M/Q_Mr \right)$.

We now take into acount the curvature of the
wire. With decreasing ring radius the angle between
neighboring spins in the domain wall shrinks. 
This leads to an additional contribution to the exchange energy of the wall
when its magnetization points out of the wire plane.

To calculate the new exchange energy we change the spin basis to Cartesian
coordinates. To distinguish the spin basis from the basis in space we
introduce the coordinates
$\chi = \cos \theta$, $\eta = \sin \theta \cos \phi$, and $\zeta =
\sin \theta \sin \phi$. 
Moving along the wire the magnetization performs a rotation in $-\theta$
direction due to the domain wall as well as a rotation around the $\zeta$ axis due to the
curvature. For small rotations $\Delta \theta$ and $\Delta \beta$ the
Cartesian coordinates are given by 
\begin{equation}
\begin{split}
\chi & = \cos(\Delta \beta) \cos(\theta + \Delta \theta) - \sin(\Delta \beta)
\sin(\theta + \Delta \theta) \cos(\phi)\\
\eta & = \cos(\Delta \beta) \sin(\theta + \Delta \theta) \cos(\phi) +
\sin(\Delta \beta) \cos(\theta + \Delta \theta)\\
\zeta & = \sin(\theta + \Delta \theta) \sin(\phi).\\
\end{split}
\label{definition_cartesian}
\end{equation}
The exchange energy density is given by
\begin{equation}
W_{\mbox{ex}} = A \left[ \left( \frac{\partial \chi}{\partial x} \right)^2 +
\left( \frac{\partial \eta}{\partial x} \right)^2 + \left( \frac{\partial
\zeta}{\partial x} \right)^2 \right].
\end{equation}
From Eq. (\ref{definition_cartesian}) we obtain
\begin{equation}
\begin{split}
W_{\mbox{ex}} & = A \left( \frac{\partial \theta}{\partial x} \right)^2 + 2 A \frac{\partial
\theta}{\partial x} \frac{1}{r} \cos \phi \\
 & - A \frac{1}{r^2} \sin^2 \theta  \sin^2 \phi  + \frac{A}{r^2}.\\
\end{split}
\end{equation}
The first term is equal to the exchange energy density of the straight wire. The
last term is constant and does not depend on the
magnetization. In the approximation for small $\phi$ the other two terms can be
rewritten
\begin{equation}
\begin{split}
\Delta W_{\mbox{ex}} & = A \frac{\partial \theta}{\partial x} \frac{1}{r}
\left( 2 - \phi^2 \right) - A \left( \frac{\phi}{r} \right)^2 \sin^2 \theta.\\
\end{split}
\end{equation}
Integration leads to the contribution
\begin{equation}
\int dV \; \Delta W_{\mbox{ex}} = \left( \frac{A S \pi}{r} - \frac{A S 2 \lambda}{r^2} \right) \phi^2 - \frac{2 A S \pi}{r}
\label{curved_wire_delta_exchange}
\end{equation}
of the curvature to the anisotropy energy.
The last term is a constant which does not depend on $X$ or $\phi$.
The perpendicular anisotropy energy can be written as
\begin{equation}
\int dV \; W_{\mbox{a} \perp} = \int dV K_{\perp} \sin^2 \theta \sin^2 \phi = K_{\perp} S 2 \lambda \phi^2.
\label{curved_wire_perpendicular_anisotropy}
\end{equation}
Comparing Eqs. (\ref{curved_wire_delta_exchange}) and
(\ref{curved_wire_perpendicular_anisotropy}) one can see that the additional
exchange energy due to the curvature can be included into the
perpendicular anisotropy by defining an effective anisotropy constant
\begin{equation}
K_{\perp \mbox{eff}} = K_{\perp} + \frac{A \pi}{2 \lambda r} - \frac{A}{r^2}.
\label{perp_anisotropy_k_eff}
\end{equation}

\section{Numerical Calculations}
\label{Numeric Calculations}

To check the applicability of the approximations made in our analytical model,
i.e. the form invariance of the domain wall at small displacements, we have
performed micromagnetic simulations. We have modelled current induced domain-wall
oscillations in curved nanowires as described in Sec. \ref{Model}.
The current contacts are arranged under an angle of
$90^\circ$ to have sufficient distance to the domain wall as well as to the
ends of the wire (see Fig. \ref{model}).

We 
extended the implementation of the Landau-Lifshitz-Gilbert-equation in the
Object Oriented Micro Magnetic Framework
(OOMMF) \cite{OOMMF} by the additional current-dependent terms of 
Eq.~(\ref{LLG-Zang-Li2}) and implemented Runge-Kutta and
Adams-Bashforth-Moulton algorithms of higher order to speed up the calculations.
The calculations presented here have been performed using the explicit embedded
Runge-Kutta 5(4) algorithm by Cash and Karp. \cite{Cash90}
The current density is calculated by locally solving Ohm's law, thus taking
the curvature of the wire and the contacts into account. 
For the spatial
discretization a cell size of 1~nm in x- and y-direction and 10~nm in z-direction
was chosen. Numerical calculations were performed for radii of 45~nm, 55~nm,
65~nm, 70~nm, 85~nm, and
95~nm with different polarized current densities $j_p = jP$. We use the material parameters of permalloy, i.e. the exchange constant
$A = 13 \cdot 10^{-12} \mbox{ J}/\mbox{m}$ and the saturation magnetization
$M_s = 8 \cdot 10^5 \mbox{ A}/\mbox{m}$. All wires have a quadratical
cross section $S = wt = 100\mbox{ nm}^2$. The applied field in $y$-direction was chosen to be $125 \mbox{ mT}$ to increase the resonance frequency,
see Eq. (\ref{resonance_frequency}), and
thus to reduce the simulation time necessary for the domain wall to perform
several oscillations. Due to the small width of the wire this high field
has virtually no effect on the ground state ($H = 0$) of the magnetization. In the ground state we
obtain a domain-wall width of $\lambda=9.25\mbox{ nm}$. The difference in the magnetization orientation $\theta$ between the analytical description of the N\'{e}el wall and the micromagnetic ground state in the curved wire is less than $5^\circ$.

\begin{figure}
  \includegraphics[angle = 270, width = 1.0\columnwidth]{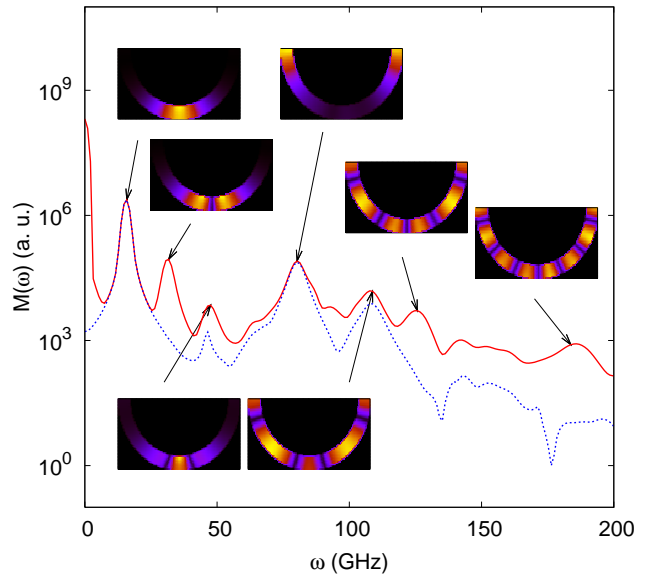}
  \caption{(Color online) Fourier transform $M(\omega)$ of the simulated magnetization $M_x(t)$ in a curved nanowire with radius $r =
45~\mbox{nm}$ and Gilbert damping parameter $\alpha = 0.05$. The wire is excited with a magnetic $\delta$-pulse. The lines
show the spatially resolved (solid line) and the integral 
response (dashed line). The insets show the spatially resolved
discrete Fourier transforms for seven selected eigenfrequencies.\label{Eigenmodes}}
\end{figure}

We have determined the eigenmodes of the magnetization in the wire by applying a magnetic $\delta$-pulse in z-direction (see Fig. \ref{model}), thus exciting all frequencies
with equal amplitude. To mimic an applied current, the
magnetic field pulse has been chosen to point in $z$-direction 
so that the torque of the 
field points in the same direction as the torque of the applied current [see
Eq. (\ref{LLG-Zang-Li2})]. After this excitation the system performs 
damped free oscillations. The eigenmodes of the wire are found by
spatially resolved discrete Fourier transformation (see Fig.
\ref{Eigenmodes}).
 \cite{McMichaelJAP2005,BayerPRB2006}
The higher harmonics and the standing spin waves in the wire
are neglected in the analytical description. The resonance of the ground mode is observed at a frequency of $\Omega = 15.7$ GHz. The higher modes are also indicated in Fig. \ref{Eigenmodes}. However in the following we focus on the ground mode. 

\begin{figure}
  \includegraphics[angle = 270, width =
1.0\columnwidth]{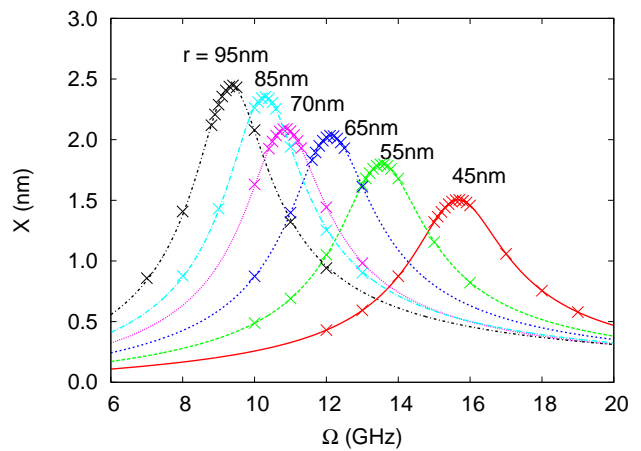}
  \caption{(Color online) Amplitude of the domain-wall displacement versus frequency of the applied current for different radii $r$. The Gilbert damping $\alpha = 0.05$,
the ratio of the exchange and spin-flip relaxation time $\xi = 0.01$, and the polarized current density $j_p = 10^{11} \mbox{A}/\mbox{m}^2$ are fixed. The crosses denote numerical values while the lines are fits with the analytical result of Eq. (\ref{sol_X(t)_phi(t)}).
\label{harmonic_fit_r}}
\end{figure}
\begin{figure}
  \includegraphics[angle = 270, width =
1.0\columnwidth]{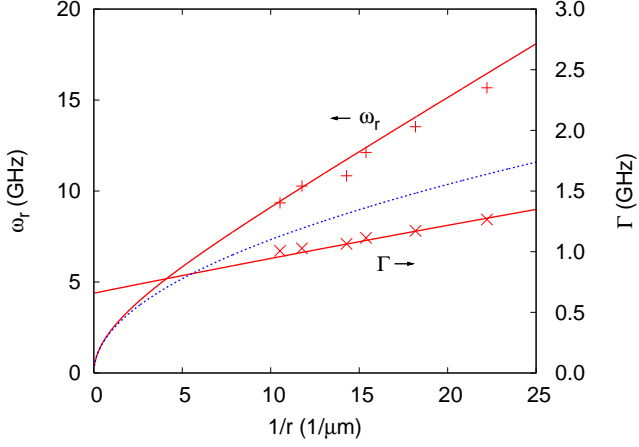}
  \caption{(Color online) Resonance frequency $\omega_r$ and damping constant $\Gamma$ versus reciprocal ring radius. Shown are the values determined from the fits in
Fig. \ref{harmonic_fit_r} (data points) and the analytical values (solid lines). The dashed line indicates the behavior of the resonance frequency as expected from the phenomenological model of Saitoh et al.\cite{Saitoh04} \label{omega_gamma_r}}
\end{figure}
\begin{figure}
  \includegraphics[angle = 270, width =
1.0\columnwidth]{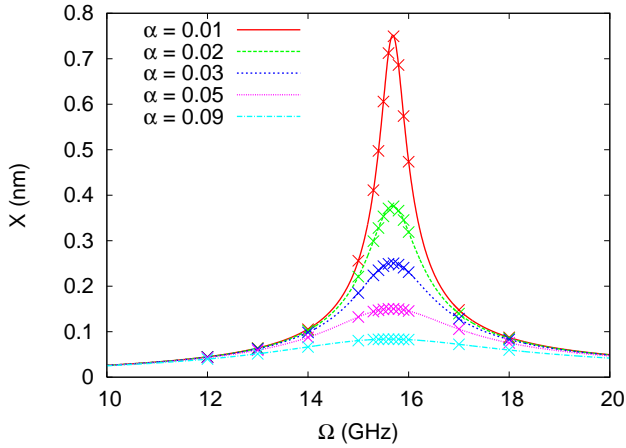}
  \caption{(Color online) Amplitude of the domain-wall displacement versus frequency of the applied current for different Gilbert damping parameters $\alpha$. The ring radius $r=45$~nm, the ratio of the exchange and spin-flip relaxation time $\xi = 0.01$, and the polarized current density $j_p = 10^{10}~
\mbox{A}/\mbox{m}^2$ are fixed. The crosses denote numerical values while the lines are fits with the analytical result of Eq. (\ref{sol_X(t)_phi(t)}).
\label{harmonic_fit_alpha}}
\end{figure}
\begin{figure}
  \includegraphics[angle = 270, width =
1.0\columnwidth]{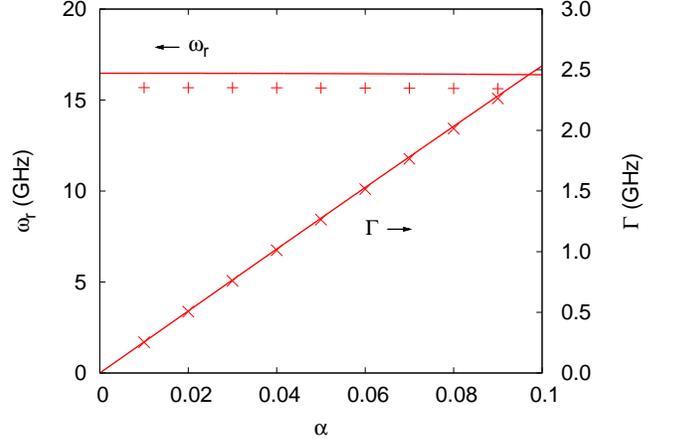}
  \caption{(Color online) Resonance frequency $\omega_r$ and damping constant $\Gamma$ versus Gilbert damping parameter $\alpha$. Shown are the values determined from the fits in Fig. \ref{harmonic_fit_alpha} (data points) and the analytical values (solid lines).
\label{omega_gamma_alpha}}
\end{figure}
\begin{figure}
  \includegraphics[angle = 270, width = 1.0\columnwidth]{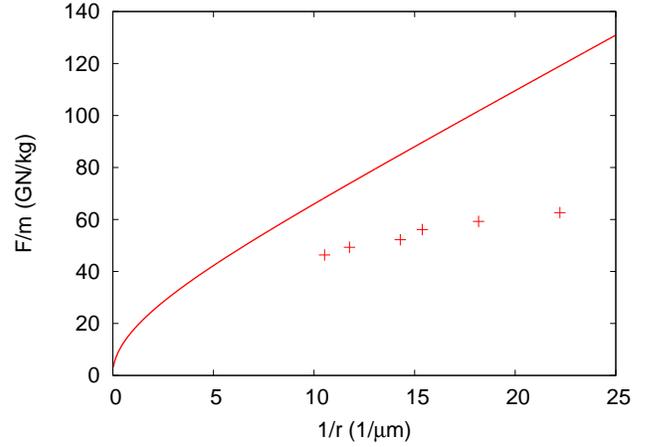}
  \caption{(Color online) Force per wall mass at the resonance frequency versus reciprocal ring radius. Shown are the numerical values (crosses) and the analytical values (line). The polarized current density is $j_p = 10^{11}~ \mbox{A}/\mbox{m}^2$. \label{force_r}}
\end{figure}
\begin{figure}
  \includegraphics[angle = 270, width =
1.0\columnwidth]{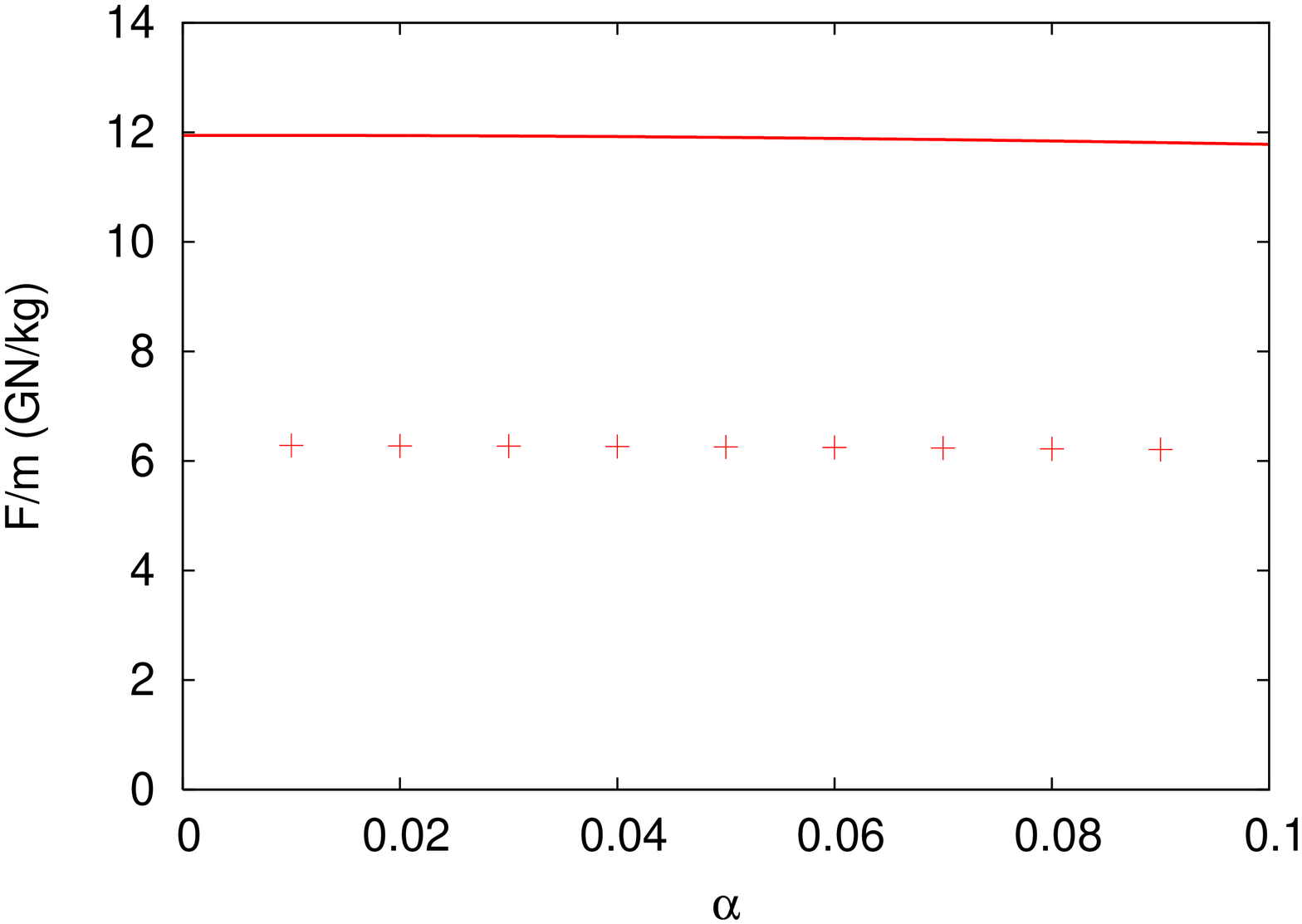}
  \caption{(Color online) Force per wall mass at the resonance frequency versus Gilbert damping parameter. Shown are the numerical (crosses) and the analytical values (line). The polarized current density is $j_p = 10^{11}~\mbox{A}/\mbox{m}^2$. \label{force_alpha}}
\end{figure}

We simulated an alternating current with frequencies close to the
resonance 
frequency of the domain wall for different radii $r$ and Gilbert damping
parameters $\alpha$. 
Figure \ref{harmonic_fit_r} shows the numerically obtained
amplitudes for different radii at fixed $\alpha = 0.05$ and $\xi = 0.01$. For each radius
the position and the width of the resonance curve have been fitted to the analytical
model, 
Eq.~(\ref{sol_X(t)_phi(t)}), to determine the parameters  $F(r)$,
$\omega_r(r)$, and $\Gamma(r)$. 
Note that all resonance curves are in excellent agreement with the harmonic-oscillator model. 
The frequencies $\omega_r(r)$ and the damping constants $\Gamma(r)$ have been summarized
in Fig.~\ref{omega_gamma_r} where they are compared to the analytical
expressions in Eq.~(\ref{oscillator_damping}) and Eq.~(\ref{resonance_frequency}).
The results coincide if we assume
$K_{\perp \mbox{eff}} = K_{\perp} + \frac{A \pi}{2 \lambda r} - \frac{A}{r^2}$
with $ K_{\perp} = 60000 \mbox{ J}/\mbox{m}^3$
for the perpendicular anisotropy [see Eq. (\ref{perp_anisotropy_k_eff})].
The dependence of the resonance frequency $\omega_r$ on the radius $r$ according to the phenomenological model of Saitoh et al. \cite{Saitoh04} is also shown.
It is visible from Fig. \ref{omega_gamma_r} that the analytical model
and the phenomenological oscillator model yield the same eigenfrequencies in the limit of a straight wire ($r >> 1~\mu\mbox{m}$). For smaller radii the phenomenological model gives eigenfrequencies
which are 
significantly lower than the ones of the numerical calculations. Our analytical model including the geometrical corrections fits the numerical data very well.

Figures \ref{harmonic_fit_alpha} and \ref{omega_gamma_alpha} show the
corresponding data  
for a ring with a radius of 45~nm and different values of the
Gilbert damping parameter $\alpha$. The analytical solutions are calculated with no
free fit parameter. While the data points for the damping constant
$\Gamma(\alpha)$ coincide with the analytical result, small deviations
occur in the resonance frequency $\omega_r(\alpha)$.
These deviations can be attributed to the finite cell size in our simulations.

In Figs. \ref{force_r} and \ref{force_alpha} the values for the fit parameter
$F(\alpha, r)$ are compared with the analytical result. The analytical values
exceed the numerically obtained parameters by up to a factor of two.
This difference has several reasons. In Sec. \ref{Analytic Calculations}
we assumed that the ground mode can be described by the motion
of the center of the
wall $X$ and the magnetization angle $\phi$. This neglects
spin-wave excitations and higher wall modes. Calculating the mode spectrum  excited with a single driving frequency
revealed a strong coupling between the ground mode and higher modes. This coupling
is enhanced for small radii. Therefore, the force is distributed over several
modes, thus decreasing the amplitude of the ground mode. Moreover, in wires
with small radii the current 
distribution is very inhomogeneous with a higher current density at their inner
edge. We expect that this leads to an additional deformation of
the N\'{e}el wall. Another aspect is the finite cell size. In the numerical calculations the curved surface has been approximated with rectangular prisms. The resulting kinks in the wire wall have a measurable effect on the domain-wall motion similar to surface roughness.

\section{Relation to Experiment}
\label{Realaton to Experiment}
\begin{figure}
  \includegraphics[angle = 270, width =
1.0\columnwidth]{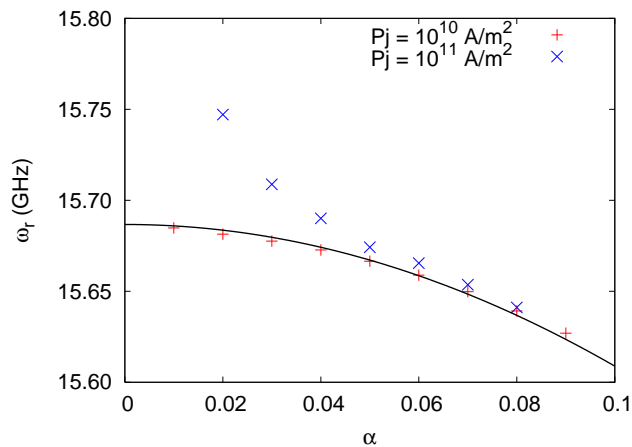}
  \caption{(Color online) Resonance frequency $\omega_r$ versus Gilbert damping parameter $\alpha$. The data points are the numerical values obtained for a wire with radius 45~nm and two different densities of the polarized current $j_p$. The line is a fit according to the analytical result $\omega_r = C/\sqrt{1+\alpha^2}$ from Eq. (\ref{resonance_frequency}) with the fit parameter $C$.
\label{omega_alpha}}
\end{figure}
In the analytical calculations we assume the linear approximations $\sin (X/r) \approx X/r$ and $\sin \phi \approx \phi$.
Nonlinearities cause the deviation of the resonance frequency in
Fig.~\ref{omega_alpha} from the analytical form at the current density
$j_p = 10^{11} \mbox{ A}/\mbox{m}^2$ and
Gilbert damping parameters below 0.05. Note that these non-linearities are
small ($<$ 0.5\%). The current densities at which non-linearities become important strongly depend
on the geometry of the wire. Our analytical model allows to derive them for typical experimental parameters. In the rest of this section we will
assume that the damping constant $\Gamma$ is not field-dependent
($\frac{\lambda H}{2r} << \frac{K_{\perp}}{\mu_0 M_s}$) and that the squared Gilbert damping parameter is small
($\alpha^2 << 1$). These assumptions usually hold in experiments because the domain-wall width $\lambda$
is small compared to the radius $r$
and the usual values of the damping parameter $\alpha$ are lower than 0.1. We can express all terms in Eq. (\ref{force due to current}) with the expressions
for $\Gamma$ and $\omega_r$ in Eqs. (\ref{oscillator_damping}) and
(\ref{resonance_frequency}) when we assume that the ratio of exchange and spin-flip relaxation time $\xi$ is comparable or less than the Gilbert damping parameter
$\alpha$. In the case of a non-critically damped oscillation
($\omega_r > \Gamma$) the oscillation becomes nonlinear if the current density
is approximately
\begin{equation}
j = \min \left( \frac{\Gamma r}{4 b_j }, \frac{\Gamma^2 \lambda}{2 \alpha \omega_r b_j } \right).
\end{equation}

The experimental current densities which 
Saitoh et al. \cite{Saitoh04} applied on a wire
with cross section $S = 3150~\mbox{nm}^2$ and radius $r = 50~ \mu \mbox{m}$
are well below this current density. They have determined a
domain-wall width $\lambda = 70~\mbox{nm}$, a domain-wall mass $m = (6.55 \pm 0.06) \cdot 10^{-23} \mbox{ kg}$, and a domain-wall relaxation time $\tau = \frac{1}{2 \Gamma} = (1.4\pm 0.2) \cdot 10^{-8} \mbox{ s}$.
To compare these findings with our analytical model we have to make an
assumption for the effective anisotropy constant $K_{\perp \mbox{eff}}$. Assuming the value $K_{\perp \mbox{eff}} = 60000
\frac{\mbox{J}}{\mbox{m}^3} + \frac{A \pi}{2 \lambda r} - \frac{A}{r^2}$ from the
fit in Sec. \ref{Numeric Calculations},
Eq. (\ref{wall_mass}) leads to the mass $m = 1.55 \cdot 10^{-23} \mbox{ kg}$. This mass is
comparable to the experimental value, despite the uncertainty of $K_{\perp \mbox{eff}}$
due to the different wire dimensions.

As mentioned in Sec.~\ref{Analytic Calculations} the analytical calculations
lead to relations between the micromagnetic material parameters and the
parameters of the harmonic oscillator. These can be used to experimentally
determine the Gilbert damping parameter $\alpha$ from the experimental data. 
From Eqs. (\ref{oscillator_damping}), (\ref{resonance_frequency}), and
(\ref{wall_mass}) one can derive the relation
\begin{equation}
\alpha = \frac{2 \Gamma \gamma H \lambda}{\omega_r^2 r} = \frac{m \Gamma \gamma
\lambda}{\mu_0 M_s S}.
\end{equation}
With the domain-wall mass and the domain-wall relaxation time of Saitoh's experiment
we get a Gilbert damping parameter of $\alpha = 0.0114 \pm 0.0017$. This value agrees quite well with the
experimental values of Nibarger \cite{Nibarger03} and Schneider \cite{Schneider05} which range from 0.008 to 0.017 for film thicknesses
between 10~nm and 93~nm.

\section{Conclusion}
\label{Conclusion}
The current-induced motion of a domain wall in thin curved nanowires has been
investigated. A harmonic-oscillator model which so far had only been introduced phenomenologically is
derived from the LLG equations extended by
the spin torque according to Zhang and Li. \cite{ZhangLiPRL2004} This derivation relates
micromagnetic material parameters to the characteristic quantities describing
the oscillating domain wall under the influence of an alternating driving
current. It is shown that the
dipole moment of the wall as well as the curvature of the wire have an important
influence on the resonance frequency and damping constant of the oscillation. The domain
wall can be seen as a quasi particle in a parabolic potential well which is acted upon
by a current-induced force. The phase and magnitude of the force depend on the
frequency of the current. The analytical results have been compared to numerical
simulations. They agree very well.
Our analytical solution suggests new methods to determine material parameters
which are otherwise difficult to measure, e.g.,
the non-adiabatic term of the spin torque can be determined
from the phase shift between the applied current and the overall
magnetization.
Moreover, the Gilbert damping parameter $\alpha$ and the domain-wall mass $m$ follow from a
measurement of the resonance frequency $\omega_r$ and
the damping constant $\Gamma$ of the oscillations.

\begin{acknowledgments}
The authors thank S. S. P. Parkin for sharing his results
prior to publication. We appreciate fruitful discussions with
U. Gummich.
Financial support by the Deutsche Forschungsgemeinschaft via SFB 668 "Magnetismus vom Einzelatom zur Nanostruktur" and via Graduiertenkolleg 1286 "Functional metal-semiconductor hybrid systems" is gratefully acknowledged.
\end{acknowledgments}


\begin{thebibliography}{48}
\expandafter\ifx\csname natexlab\endcsname\relax\def\natexlab#1{#1}\fi
\expandafter\ifx\csname bibnamefont\endcsname\relax
  \def\bibnamefont#1{#1}\fi
\expandafter\ifx\csname bibfnamefont\endcsname\relax
  \def\bibfnamefont#1{#1}\fi
\expandafter\ifx\csname citenamefont\endcsname\relax
  \def\citenamefont#1{#1}\fi
\expandafter\ifx\csname url\endcsname\relax
  \def\url#1{\texttt{#1}}\fi
\expandafter\ifx\csname urlprefix\endcsname\relax\def\urlprefix{URL }\fi
\providecommand{\bibinfo}[2]{#2}
\providecommand{\eprint}[2][]{\url{#2}}

\bibitem[{\citenamefont{Thiele}(1974)}]{Thiele1974}
\bibinfo{author}{\bibfnamefont{A.~A.} \bibnamefont{Thiele}},
  \bibinfo{journal}{J. Appl. Phys.} \textbf{\bibinfo{volume}{45}},
  \bibinfo{pages}{377} (\bibinfo{year}{1974}).

\bibitem[{\citenamefont{Schreyer and Walker}(1974)}]{Walker1974}
\bibinfo{author}{\bibfnamefont{N.~L.} \bibnamefont{Schreyer}} \bibnamefont{and}
  \bibinfo{author}{\bibfnamefont{L.~R.} \bibnamefont{Walker}},
  \bibinfo{journal}{J. Appl. Phys.} \textbf{\bibinfo{volume}{45}},
  \bibinfo{pages}{5406} (\bibinfo{year}{1974}).

\bibitem[{\citenamefont{Slonczewski}(1996)}]{Slonczewski1996}
\bibinfo{author}{\bibfnamefont{J.}~\bibnamefont{Slonczewski}},
  \bibinfo{journal}{J. Magn. Magn. Mater.} \textbf{\bibinfo{volume}{159}},
  \bibinfo{pages}{L1} (\bibinfo{year}{1996}).

\bibitem[{\citenamefont{Berger}(1996)}]{Berger1996}
\bibinfo{author}{\bibfnamefont{L.}~\bibnamefont{Berger}},
  \bibinfo{journal}{Phys. Rev. B} \textbf{\bibinfo{volume}{54}},
  \bibinfo{pages}{9353} (\bibinfo{year}{1996}).

\bibitem[{\citenamefont{Myers et~al.}(1999)\citenamefont{Myers, Ralph, Katine,
  Louie, and Buhrman}}]{Myers1999}
\bibinfo{author}{\bibfnamefont{E.~B.} \bibnamefont{Myers}},
  \bibinfo{author}{\bibfnamefont{D.~C.} \bibnamefont{Ralph}},
  \bibinfo{author}{\bibfnamefont{J.~A.} \bibnamefont{Katine}},
  \bibinfo{author}{\bibfnamefont{R.~N.} \bibnamefont{Louie}}, \bibnamefont{and}
  \bibinfo{author}{\bibfnamefont{R.~A.} \bibnamefont{Buhrman}},
  \bibinfo{journal}{Science} \textbf{\bibinfo{volume}{285}},
  \bibinfo{pages}{867} (\bibinfo{year}{1999}).

\bibitem[{\citenamefont{Grollier et~al.}(2003)\citenamefont{Grollier, Boulenc,
  Cros, Hamzi, Vaur, Fert, and Faini}}]{GrollierAPL2003}
\bibinfo{author}{\bibfnamefont{J.}~\bibnamefont{Grollier}},
  \bibinfo{author}{\bibfnamefont{P.}~\bibnamefont{Boulenc}},
  \bibinfo{author}{\bibfnamefont{V.}~\bibnamefont{Cros}},
  \bibinfo{author}{\bibfnamefont{A.}~\bibnamefont{Hamzi}},
  \bibinfo{author}{\bibfnamefont{A.}~\bibnamefont{Vaur}},
  \bibinfo{author}{\bibfnamefont{A.}~\bibnamefont{Fert}}, \bibnamefont{and}
  \bibinfo{author}{\bibfnamefont{G.}~\bibnamefont{Faini}},
  \bibinfo{journal}{Appl. Phys. Lett.} \textbf{\bibinfo{volume}{83}},
  \bibinfo{pages}{509} (\bibinfo{year}{2003}).

\bibitem[{\citenamefont{Yamaguchi et~al.}(2004)\citenamefont{Yamaguchi, Ono,
  Nasu, Miyake, Mibu, and Shinjo}}]{Yamaguchi2004}
\bibinfo{author}{\bibfnamefont{A.}~\bibnamefont{Yamaguchi}},
  \bibinfo{author}{\bibfnamefont{T.}~\bibnamefont{Ono}},
  \bibinfo{author}{\bibfnamefont{S.}~\bibnamefont{Nasu}},
  \bibinfo{author}{\bibfnamefont{K.}~\bibnamefont{Miyake}},
  \bibinfo{author}{\bibfnamefont{K.}~\bibnamefont{Mibu}}, \bibnamefont{and}
  \bibinfo{author}{\bibfnamefont{T.}~\bibnamefont{Shinjo}},
  \bibinfo{journal}{Phys. Rev. Lett.} \textbf{\bibinfo{volume}{92}},
  \bibinfo{pages}{077205} (\bibinfo{year}{2004}).

\bibitem[{\citenamefont{Koo et~al.}(2002)\citenamefont{Koo, Krafft, and
  Gomez}}]{Koo2002}
\bibinfo{author}{\bibfnamefont{H.}~\bibnamefont{Koo}},
  \bibinfo{author}{\bibfnamefont{C.}~\bibnamefont{Krafft}}, \bibnamefont{and}
  \bibinfo{author}{\bibfnamefont{R.~D.} \bibnamefont{Gomez}},
  \bibinfo{journal}{Appl. Phys. Lett.} \textbf{\bibinfo{volume}{81}},
  \bibinfo{pages}{862} (\bibinfo{year}{2002}).

\bibitem[{\citenamefont{Saitoh et~al.}(2004)\citenamefont{Saitoh, Miyajima,
  Yamaoka, and Tatara}}]{Saitoh04}
\bibinfo{author}{\bibfnamefont{E.}~\bibnamefont{Saitoh}},
  \bibinfo{author}{\bibfnamefont{H.}~\bibnamefont{Miyajima}},
  \bibinfo{author}{\bibfnamefont{T.}~\bibnamefont{Yamaoka}}, \bibnamefont{and}
  \bibinfo{author}{\bibfnamefont{G.}~\bibnamefont{Tatara}},
  \bibinfo{journal}{Nature} \textbf{\bibinfo{volume}{432}},
  \bibinfo{pages}{203} (\bibinfo{year}{2004}).

\bibitem[{\citenamefont{Klaeui et~al.}(2003)\citenamefont{Klaeui, Vaz, Bland,
  Wernsdorfer, Faini, Cambril, and Heyderman}}]{KlaeuiAPL2003}
\bibinfo{author}{\bibfnamefont{M.}~\bibnamefont{Klaeui}},
  \bibinfo{author}{\bibfnamefont{C.~A.~F.} \bibnamefont{Vaz}},
  \bibinfo{author}{\bibfnamefont{J.~A.~C.} \bibnamefont{Bland}},
  \bibinfo{author}{\bibfnamefont{W.}~\bibnamefont{Wernsdorfer}},
  \bibinfo{author}{\bibfnamefont{G.}~\bibnamefont{Faini}},
  \bibinfo{author}{\bibfnamefont{E.}~\bibnamefont{Cambril}}, \bibnamefont{and}
  \bibinfo{author}{\bibfnamefont{L.~J.} \bibnamefont{Heyderman}},
  \bibinfo{journal}{Appl. Phys. Lett.} \textbf{\bibinfo{volume}{83}},
  \bibinfo{pages}{105} (\bibinfo{year}{2003}).

\bibitem[{\citenamefont{Vernier et~al.}(2004)\citenamefont{Vernier, Allwood,
  Atkinson, Cooke, and Cowburn}}]{VernierEPL2004}
\bibinfo{author}{\bibfnamefont{N.}~\bibnamefont{Vernier}},
  \bibinfo{author}{\bibfnamefont{D.}~\bibnamefont{Allwood}},
  \bibinfo{author}{\bibfnamefont{D.}~\bibnamefont{Atkinson}},
  \bibinfo{author}{\bibfnamefont{M.}~\bibnamefont{Cooke}}, \bibnamefont{and}
  \bibinfo{author}{\bibfnamefont{R.}~\bibnamefont{Cowburn}},
  \bibinfo{journal}{Europhys. Lett.} \textbf{\bibinfo{volume}{65}},
  \bibinfo{pages}{526} (\bibinfo{year}{2004}).

\bibitem[{\citenamefont{Katine et~al.}(2000)\citenamefont{Katine, Albert,
  Buhrman, Myers, and Ralph}}]{KatinePRL2000}
\bibinfo{author}{\bibfnamefont{J.~A.} \bibnamefont{Katine}},
  \bibinfo{author}{\bibfnamefont{F.~J.} \bibnamefont{Albert}},
  \bibinfo{author}{\bibfnamefont{R.~A.} \bibnamefont{Buhrman}},
  \bibinfo{author}{\bibfnamefont{E.~B.} \bibnamefont{Myers}}, \bibnamefont{and}
  \bibinfo{author}{\bibfnamefont{D.~C.} \bibnamefont{Ralph}},
  \bibinfo{journal}{Phys. Rev. Lett.} \textbf{\bibinfo{volume}{84}},
  \bibinfo{pages}{3149} (\bibinfo{year}{2000}).

\bibitem[{\citenamefont{Krivorotov et~al.}(2005)\citenamefont{Krivorotov,
  Emley, Sankey, Kiselev, Ralph, and Buhrman}}]{KrivorotovScience2005}
\bibinfo{author}{\bibfnamefont{I.~N.} \bibnamefont{Krivorotov}},
  \bibinfo{author}{\bibfnamefont{N.~C.} \bibnamefont{Emley}},
  \bibinfo{author}{\bibfnamefont{J.~C.} \bibnamefont{Sankey}},
  \bibinfo{author}{\bibfnamefont{S.~I.} \bibnamefont{Kiselev}},
  \bibinfo{author}{\bibfnamefont{D.~C.} \bibnamefont{Ralph}}, \bibnamefont{and}
  \bibinfo{author}{\bibfnamefont{R.~A.} \bibnamefont{Buhrman}},
  \bibinfo{journal}{Science} \textbf{\bibinfo{volume}{307}},
  \bibinfo{pages}{228} (\bibinfo{year}{2005}).

\bibitem[{\citenamefont{Parkin}(2004)}]{ParkinPatent2004}
\bibinfo{author}{\bibfnamefont{S.~S.~P.} \bibnamefont{Parkin}},
  \bibinfo{journal}{US Patent} \textbf{\bibinfo{volume}{309}},
  \bibinfo{pages}{6,834,005} (\bibinfo{year}{2004}).

\bibitem[{\citenamefont{Allwood et~al.}(2002)\citenamefont{Allwood, Xiong,
  Cooke, Faulkner, Atkinson, Vernier, and Cowburn}}]{AllwoodScience2002}
\bibinfo{author}{\bibfnamefont{D.~A.} \bibnamefont{Allwood}},
  \bibinfo{author}{\bibfnamefont{G.}~\bibnamefont{Xiong}},
  \bibinfo{author}{\bibfnamefont{M.~D.} \bibnamefont{Cooke}},
  \bibinfo{author}{\bibfnamefont{C.~C.} \bibnamefont{Faulkner}},
  \bibinfo{author}{\bibfnamefont{D.}~\bibnamefont{Atkinson}},
  \bibinfo{author}{\bibfnamefont{N.}~\bibnamefont{Vernier}}, \bibnamefont{and}
  \bibinfo{author}{\bibfnamefont{R.~P.} \bibnamefont{Cowburn}},
  \bibinfo{journal}{Science} \textbf{\bibinfo{volume}{296}},
  \bibinfo{pages}{2003} (\bibinfo{year}{2002}).

\bibitem[{\citenamefont{Allwood et~al.}(2005)\citenamefont{Allwood, Xiong,
  Faulkner, Atkinson, Petit, and Cowburn}}]{AllwoodScience2005}
\bibinfo{author}{\bibfnamefont{D.~A.} \bibnamefont{Allwood}},
  \bibinfo{author}{\bibfnamefont{G.}~\bibnamefont{Xiong}},
  \bibinfo{author}{\bibfnamefont{C.~C.} \bibnamefont{Faulkner}},
  \bibinfo{author}{\bibfnamefont{D.}~\bibnamefont{Atkinson}},
  \bibinfo{author}{\bibfnamefont{D.}~\bibnamefont{Petit}}, \bibnamefont{and}
  \bibinfo{author}{\bibfnamefont{R.~P.} \bibnamefont{Cowburn}},
  \bibinfo{journal}{Science} \textbf{\bibinfo{volume}{309}},
  \bibinfo{pages}{1688} (\bibinfo{year}{2005}).

\bibitem[{\citenamefont{Cowburn}(2004)}]{CowburnPatent2004}
\bibinfo{author}{\bibfnamefont{R.~P.} \bibnamefont{Cowburn}},
  \bibinfo{journal}{Patent Application} \textbf{\bibinfo{volume}{309}},
  \bibinfo{pages}{WO002004077451A1} (\bibinfo{year}{2004}).

\bibitem[{\citenamefont{Tatara and Kohno}(2004)}]{TataraPRL2004}
\bibinfo{author}{\bibfnamefont{G.}~\bibnamefont{Tatara}} \bibnamefont{and}
  \bibinfo{author}{\bibfnamefont{H.}~\bibnamefont{Kohno}},
  \bibinfo{journal}{Phys. Rev. Lett.} \textbf{\bibinfo{volume}{92}},
  \bibinfo{pages}{086601} (\bibinfo{year}{2004}).

\bibitem[{\citenamefont{Thiaville et~al.}(2005)\citenamefont{Thiaville,
  Nakatani, Miltat, and Suzuki}}]{ThiavilleEPL2005}
\bibinfo{author}{\bibfnamefont{A.}~\bibnamefont{Thiaville}},
  \bibinfo{author}{\bibfnamefont{Y.}~\bibnamefont{Nakatani}},
  \bibinfo{author}{\bibfnamefont{J.}~\bibnamefont{Miltat}}, \bibnamefont{and}
  \bibinfo{author}{\bibfnamefont{Y.}~\bibnamefont{Suzuki}},
  \bibinfo{journal}{Europhys. Lett.} \textbf{\bibinfo{volume}{69}},
  \bibinfo{pages}{990} (\bibinfo{year}{2005}).

\bibitem[{\citenamefont{Waintal and Viret}(2004)}]{WaintalEPL2004}
\bibinfo{author}{\bibfnamefont{X.}~\bibnamefont{Waintal}} \bibnamefont{and}
  \bibinfo{author}{\bibfnamefont{M.}~\bibnamefont{Viret}},
  \bibinfo{journal}{Europhys. Lett.} \textbf{\bibinfo{volume}{65}},
  \bibinfo{pages}{427} (\bibinfo{year}{2004}).

\bibitem[{\citenamefont{Zhang and Li}(2004{\natexlab{a}})}]{ZhangLiPRL2004}
\bibinfo{author}{\bibfnamefont{S.}~\bibnamefont{Zhang}} \bibnamefont{and}
  \bibinfo{author}{\bibfnamefont{Z.}~\bibnamefont{Li}}, \bibinfo{journal}{Phys.
  Rev. Lett.} \textbf{\bibinfo{volume}{93}}, \bibinfo{pages}{127204}
  (\bibinfo{year}{2004}{\natexlab{a}}).

\bibitem[{\citenamefont{Zhang and Li}(2004{\natexlab{b}})}]{ZhangLi2004a}
\bibinfo{author}{\bibfnamefont{S.}~\bibnamefont{Zhang}} \bibnamefont{and}
  \bibinfo{author}{\bibfnamefont{Z.}~\bibnamefont{Li}}, \bibinfo{journal}{Phys.
  Rev. B} \textbf{\bibinfo{volume}{70}}, \bibinfo{pages}{024417}
  (\bibinfo{year}{2004}{\natexlab{b}}).

\bibitem[{\citenamefont{Klaeui et~al.}(2005{\natexlab{a}})\citenamefont{Klaeui,
  Vaz, Bland, Wernsdorfer, Faini, Cambril, Heyderman, Nolting, and
  Ruediger}}]{KlaeuiPRL2005}
\bibinfo{author}{\bibfnamefont{M.}~\bibnamefont{Klaeui}},
  \bibinfo{author}{\bibfnamefont{C.~A.~F.} \bibnamefont{Vaz}},
  \bibinfo{author}{\bibfnamefont{J.~A.~C.} \bibnamefont{Bland}},
  \bibinfo{author}{\bibfnamefont{W.}~\bibnamefont{Wernsdorfer}},
  \bibinfo{author}{\bibfnamefont{G.}~\bibnamefont{Faini}},
  \bibinfo{author}{\bibfnamefont{E.}~\bibnamefont{Cambril}},
  \bibinfo{author}{\bibfnamefont{L.~J.} \bibnamefont{Heyderman}},
  \bibinfo{author}{\bibfnamefont{F.}~\bibnamefont{Nolting}}, \bibnamefont{and}
  \bibinfo{author}{\bibfnamefont{U.}~\bibnamefont{Ruediger}},
  \bibinfo{journal}{Phys. Rev. Lett.} \textbf{\bibinfo{volume}{94}},
  \bibinfo{pages}{106601} (\bibinfo{year}{2005}{\natexlab{a}}).

\bibitem[{\citenamefont{Ono et~al.}(1999)\citenamefont{Ono, Miyajima, Shigeto,
  Mibu, Hosoito, and Shinjo}}]{OnoScience1999}
\bibinfo{author}{\bibfnamefont{T.}~\bibnamefont{Ono}},
  \bibinfo{author}{\bibfnamefont{H.}~\bibnamefont{Miyajima}},
  \bibinfo{author}{\bibfnamefont{K.}~\bibnamefont{Shigeto}},
  \bibinfo{author}{\bibfnamefont{K.}~\bibnamefont{Mibu}},
  \bibinfo{author}{\bibfnamefont{N.}~\bibnamefont{Hosoito}}, \bibnamefont{and}
  \bibinfo{author}{\bibfnamefont{T.}~\bibnamefont{Shinjo}},
  \bibinfo{journal}{Science} \textbf{\bibinfo{volume}{284}},
  \bibinfo{pages}{468} (\bibinfo{year}{1999}).

\bibitem[{\citenamefont{Nakatani et~al.}(2003)\citenamefont{Nakatani,
  Thiaville, and Miltat}}]{ThiavilleNatureMat}
\bibinfo{author}{\bibfnamefont{Y.}~\bibnamefont{Nakatani}},
  \bibinfo{author}{\bibfnamefont{A.}~\bibnamefont{Thiaville}},
  \bibnamefont{and} \bibinfo{author}{\bibfnamefont{J.}~\bibnamefont{Miltat}},
  \bibinfo{journal}{Nature Materials} \textbf{\bibinfo{volume}{2}},
  \bibinfo{pages}{521} (\bibinfo{year}{2003}).

\bibitem[{\citenamefont{Klaeui et~al.}(2005{\natexlab{b}})\citenamefont{Klaeui,
  Jubert, Allenspach, Bischof, Bland, Faini, Ruediger, Vaz, Vila, and
  Vouille}}]{KlaeuiPRL2005a}
\bibinfo{author}{\bibfnamefont{M.}~\bibnamefont{Klaeui}},
  \bibinfo{author}{\bibfnamefont{P.-O.} \bibnamefont{Jubert}},
  \bibinfo{author}{\bibfnamefont{R.}~\bibnamefont{Allenspach}},
  \bibinfo{author}{\bibfnamefont{A.}~\bibnamefont{Bischof}},
  \bibinfo{author}{\bibfnamefont{J.~A.~C.} \bibnamefont{Bland}},
  \bibinfo{author}{\bibfnamefont{G.}~\bibnamefont{Faini}},
  \bibinfo{author}{\bibfnamefont{U.}~\bibnamefont{Ruediger}},
  \bibinfo{author}{\bibfnamefont{C.~A.~F.} \bibnamefont{Vaz}},
  \bibinfo{author}{\bibfnamefont{L.}~\bibnamefont{Vila}}, \bibnamefont{and}
  \bibinfo{author}{\bibfnamefont{C.}~\bibnamefont{Vouille}},
  \bibinfo{journal}{Phys. Rev. Lett.} \textbf{\bibinfo{volume}{95}},
  \bibinfo{pages}{026601} (\bibinfo{year}{2005}{\natexlab{b}}).

\bibitem[{\citenamefont{Yamaguchi et~al.}(2005)\citenamefont{Yamaguchi, Nasu,
  Tanigawa, Ono, Miyake, Mibu, and Shinjo}}]{YamaguchiAPL2005}
\bibinfo{author}{\bibfnamefont{A.}~\bibnamefont{Yamaguchi}},
  \bibinfo{author}{\bibfnamefont{S.}~\bibnamefont{Nasu}},
  \bibinfo{author}{\bibfnamefont{H.}~\bibnamefont{Tanigawa}},
  \bibinfo{author}{\bibfnamefont{T.}~\bibnamefont{Ono}},
  \bibinfo{author}{\bibfnamefont{K.}~\bibnamefont{Miyake}},
  \bibinfo{author}{\bibfnamefont{K.}~\bibnamefont{Mibu}}, \bibnamefont{and}
  \bibinfo{author}{\bibfnamefont{T.}~\bibnamefont{Shinjo}},
  \bibinfo{journal}{Appl. Phys. Lett.} \textbf{\bibinfo{volume}{86}},
  \bibinfo{pages}{012511} (\bibinfo{year}{2005}).

\bibitem[{\citenamefont{Yamaguchi et~al.}(2006)\citenamefont{Yamaguchi, Ono,
  Nasu, Miyake, Mibu, and Shinjo}}]{YamaguchiErratum2006}
\bibinfo{author}{\bibfnamefont{A.}~\bibnamefont{Yamaguchi}},
  \bibinfo{author}{\bibfnamefont{T.}~\bibnamefont{Ono}},
  \bibinfo{author}{\bibfnamefont{S.}~\bibnamefont{Nasu}},
  \bibinfo{author}{\bibfnamefont{K.}~\bibnamefont{Miyake}},
  \bibinfo{author}{\bibfnamefont{K.}~\bibnamefont{Mibu}}, \bibnamefont{and}
  \bibinfo{author}{\bibfnamefont{T.}~\bibnamefont{Shinjo}},
  \bibinfo{journal}{Phys. Rev. Lett.} \textbf{\bibinfo{volume}{96}},
  \bibinfo{pages}{179904} (\bibinfo{year}{2006}).

\bibitem[{\citenamefont{He et~al.}(2006)\citenamefont{He, Li, and
  Zhang}}]{HePRB2006}
\bibinfo{author}{\bibfnamefont{J.}~\bibnamefont{He}},
  \bibinfo{author}{\bibfnamefont{Z.}~\bibnamefont{Li}}, \bibnamefont{and}
  \bibinfo{author}{\bibfnamefont{S.}~\bibnamefont{Zhang}},
  \bibinfo{journal}{Phys. Rev. B} \textbf{\bibinfo{volume}{73}},
  \bibinfo{pages}{184408} (\bibinfo{year}{2006}).

\bibitem[{\citenamefont{Klaeui et~al.}(2006)\citenamefont{Klaeui, Laufenberg,
  Heyne, Backes, Ruediger, Vaz, Bland, Heyderman, Cherifi, Locatelli
  et~al.}}]{KlaeuiAPL2006}
\bibinfo{author}{\bibfnamefont{M.}~\bibnamefont{Klaeui}},
  \bibinfo{author}{\bibfnamefont{M.}~\bibnamefont{Laufenberg}},
  \bibinfo{author}{\bibfnamefont{L.}~\bibnamefont{Heyne}},
  \bibinfo{author}{\bibfnamefont{D.}~\bibnamefont{Backes}},
  \bibinfo{author}{\bibfnamefont{U.}~\bibnamefont{Ruediger}},
  \bibinfo{author}{\bibfnamefont{C.~A.~F.} \bibnamefont{Vaz}},
  \bibinfo{author}{\bibfnamefont{J.~A.~C.} \bibnamefont{Bland}},
  \bibinfo{author}{\bibfnamefont{L.~J.} \bibnamefont{Heyderman}},
  \bibinfo{author}{\bibfnamefont{S.}~\bibnamefont{Cherifi}},
  \bibinfo{author}{\bibfnamefont{A.}~\bibnamefont{Locatelli}},
  \bibnamefont{et~al.}, \bibinfo{journal}{Appl. Phys. Lett.}
  \textbf{\bibinfo{volume}{88}}, \bibinfo{pages}{232507}
  (\bibinfo{year}{2006}).

\bibitem[{\citenamefont{Meier et~al.}()\citenamefont{Meier, Bolte, Eiselt,
  Merkt, Krueger, Pfannkuche, Kim, and Fischer}}]{ourALSdata}
\bibinfo{author}{\bibfnamefont{G.}~\bibnamefont{Meier}},
  \bibinfo{author}{\bibfnamefont{M.}~\bibnamefont{Bolte}},
  \bibinfo{author}{\bibfnamefont{R.}~\bibnamefont{Eiselt}},
  \bibinfo{author}{\bibfnamefont{U.}~\bibnamefont{Merkt}},
  \bibinfo{author}{\bibfnamefont{B.}~\bibnamefont{Krueger}},
  \bibinfo{author}{\bibfnamefont{D.}~\bibnamefont{Pfannkuche}},
  \bibinfo{author}{\bibfnamefont{D.~H.} \bibnamefont{Kim}}, \bibnamefont{and}
  \bibinfo{author}{\bibfnamefont{P.}~\bibnamefont{Fischer}},
  \bibinfo{note}{unpublished}.

\bibitem[{\citenamefont{Atkinson et~al.}(2003)\citenamefont{Atkinson, Allwood,
  Xiong, Cooke, Faulkner, and Cowburn}}]{AtkinsonNature2003}
\bibinfo{author}{\bibfnamefont{D.}~\bibnamefont{Atkinson}},
  \bibinfo{author}{\bibfnamefont{D.}~\bibnamefont{Allwood}},
  \bibinfo{author}{\bibfnamefont{G.}~\bibnamefont{Xiong}},
  \bibinfo{author}{\bibfnamefont{M.~D.} \bibnamefont{Cooke}},
  \bibinfo{author}{\bibfnamefont{C.~C.} \bibnamefont{Faulkner}},
  \bibnamefont{and} \bibinfo{author}{\bibfnamefont{R.~P.}
  \bibnamefont{Cowburn}}, \bibinfo{journal}{Nature Materials}
  \textbf{\bibinfo{volume}{2}}, \bibinfo{pages}{85} (\bibinfo{year}{2003}).

\bibitem[{\citenamefont{Hayashi et~al.}(2006)\citenamefont{Hayashi, Thomas,
  Bazaliy, Rettner, Moriya, Jiang, and Parkin}}]{Hayashi2006}
\bibinfo{author}{\bibfnamefont{M.}~\bibnamefont{Hayashi}},
  \bibinfo{author}{\bibfnamefont{L.}~\bibnamefont{Thomas}},
  \bibinfo{author}{\bibfnamefont{Y.~B.} \bibnamefont{Bazaliy}},
  \bibinfo{author}{\bibfnamefont{C.}~\bibnamefont{Rettner}},
  \bibinfo{author}{\bibfnamefont{R.}~\bibnamefont{Moriya}},
  \bibinfo{author}{\bibfnamefont{X.}~\bibnamefont{Jiang}}, \bibnamefont{and}
  \bibinfo{author}{\bibfnamefont{S.~S.~P.} \bibnamefont{Parkin}},
  \bibinfo{journal}{Phys. Rev. Lett.} \textbf{\bibinfo{volume}{96}},
  \bibinfo{pages}{197207} (\bibinfo{year}{2006}).

\bibitem[{\citenamefont{Thomas et~al.}(2006)\citenamefont{Thomas, Hayashi,
  Moriya, Jiang, Rettner, and Parkin}}]{Thomas2006}
\bibinfo{author}{\bibfnamefont{L.}~\bibnamefont{Thomas}},
  \bibinfo{author}{\bibfnamefont{M.}~\bibnamefont{Hayashi}},
  \bibinfo{author}{\bibfnamefont{R.}~\bibnamefont{Moriya}},
  \bibinfo{author}{\bibfnamefont{X.}~\bibnamefont{Jiang}},
  \bibinfo{author}{\bibfnamefont{C.}~\bibnamefont{Rettner}}, \bibnamefont{and}
  \bibinfo{author}{\bibfnamefont{S.~S.~P.} \bibnamefont{Parkin}},
  \bibinfo{journal}{Nature in press}  (\bibinfo{year}{2006}).

\bibitem[{\citenamefont{Tatara et~al.}(2005)\citenamefont{Tatara, Saitoh,
  Ichimura, and Kohno}}]{TataraAPL2005}
\bibinfo{author}{\bibfnamefont{G.}~\bibnamefont{Tatara}},
  \bibinfo{author}{\bibfnamefont{E.}~\bibnamefont{Saitoh}},
  \bibinfo{author}{\bibfnamefont{M.}~\bibnamefont{Ichimura}}, \bibnamefont{and}
  \bibinfo{author}{\bibfnamefont{H.}~\bibnamefont{Kohno}},
  \bibinfo{journal}{Appl. Phys. Lett.} \textbf{\bibinfo{volume}{86}},
  \bibinfo{pages}{232504} (\bibinfo{year}{2005}).

\bibitem[{\citenamefont{McMichael and Donahue}(1997)}]{McMichaelIEEE}
\bibinfo{author}{\bibfnamefont{R.~D.} \bibnamefont{McMichael}}
  \bibnamefont{and} \bibinfo{author}{\bibfnamefont{M.~J.}
  \bibnamefont{Donahue}}, \bibinfo{journal}{IEEE Trans. Magn.}
  \textbf{\bibinfo{volume}{33}}, \bibinfo{pages}{4167} (\bibinfo{year}{1997}).

\bibitem[{\citenamefont{Klaeui et~al.}(2004)\citenamefont{Klaeui, Vaz, Bland,
  Heyderman, Nolting, Pavlovska, Bauer, Cherifi, Heun, and
  Locatelli}}]{KlaeuiPhaseDiagram}
\bibinfo{author}{\bibfnamefont{M.}~\bibnamefont{Klaeui}},
  \bibinfo{author}{\bibfnamefont{C.~A.~F.} \bibnamefont{Vaz}},
  \bibinfo{author}{\bibfnamefont{J.~A.~C.} \bibnamefont{Bland}},
  \bibinfo{author}{\bibfnamefont{L.~J.} \bibnamefont{Heyderman}},
  \bibinfo{author}{\bibfnamefont{F.}~\bibnamefont{Nolting}},
  \bibinfo{author}{\bibfnamefont{A.}~\bibnamefont{Pavlovska}},
  \bibinfo{author}{\bibfnamefont{E.}~\bibnamefont{Bauer}},
  \bibinfo{author}{\bibfnamefont{S.}~\bibnamefont{Cherifi}},
  \bibinfo{author}{\bibfnamefont{S.}~\bibnamefont{Heun}}, \bibnamefont{and}
  \bibinfo{author}{\bibfnamefont{A.}~\bibnamefont{Locatelli}},
  \bibinfo{journal}{Appl. Phys. Lett.} \textbf{\bibinfo{volume}{85}},
  \bibinfo{pages}{5637} (\bibinfo{year}{2004}).

\bibitem[{\citenamefont{Nakatani et~al.}(2005)\citenamefont{Nakatani,
  Thiaville, and Miltat}}]{ThiavilleJMMM}
\bibinfo{author}{\bibfnamefont{Y.}~\bibnamefont{Nakatani}},
  \bibinfo{author}{\bibfnamefont{A.}~\bibnamefont{Thiaville}},
  \bibnamefont{and} \bibinfo{author}{\bibfnamefont{J.}~\bibnamefont{Miltat}},
  \bibinfo{journal}{J. Magn. Magn. Mater.} \textbf{\bibinfo{volume}{290}},
  \bibinfo{pages}{750} (\bibinfo{year}{2005}).

\bibitem[{Pre()}]{Preparation}
\bibinfo{note}{Starting with a high magnetic field the wire is saturated in
  $y$-direction. Reducing the magnetic field to a small offset the system ends
  up with a domain wall at the bottom of the wire. The wall type depends on the
  width $w$ and the thickness $t$ of the wire\cite{ThiavilleJMMM}. In the
  present study we choose a small cross section $S = wt = 100~\mbox{nm}^2$ with
  width $w = 10~\mbox{nm}$ and thickness $t = 10~\mbox{nm}$ to get a transverse
  wall.}

\bibitem[{\citenamefont{Landau and Lifshitz}(1935)}]{LLG}
\bibinfo{author}{\bibfnamefont{L.}~\bibnamefont{Landau}} \bibnamefont{and}
  \bibinfo{author}{\bibfnamefont{E.}~\bibnamefont{Lifshitz}},
  \bibinfo{journal}{Physik. Z. Sowjetunion} \textbf{\bibinfo{volume}{8}},
  \bibinfo{pages}{153} (\bibinfo{year}{1935}).

\bibitem[{Ani()}]{Anisotropy}
\bibinfo{note}{Uniaxial anisotropy can be taken into acount by replacing the
  anisotropy constants $K$ and $K_{\perp \mbox{eff}}$ with the new constants
  $K'$ and $K'_{\perp}$ given by $K' = K + K_y - K_x$ and $K'_{\perp} =
  K_{\perp \mbox{eff}} + K_z - K_y$}.

\bibitem[{\citenamefont{Dugaev et~al.}(2006)\citenamefont{Dugaev, Vieira,
  Sacramento, Barna\'{s}, Ara\'{u}jo, and Berakdar}}]{Dugaev06}
\bibinfo{author}{\bibfnamefont{V.~K.} \bibnamefont{Dugaev}},
  \bibinfo{author}{\bibfnamefont{V.~R.} \bibnamefont{Vieira}},
  \bibinfo{author}{\bibfnamefont{P.~D.} \bibnamefont{Sacramento}},
  \bibinfo{author}{\bibfnamefont{J.}~\bibnamefont{Barna\'{s}}},
  \bibinfo{author}{\bibfnamefont{M.~A.~N.} \bibnamefont{Ara\'{u}jo}},
  \bibnamefont{and} \bibinfo{author}{\bibfnamefont{J.}~\bibnamefont{Berakdar}},
  \bibinfo{journal}{Phys. Rev. B} \textbf{\bibinfo{volume}{74}},
  \bibinfo{pages}{054403} (\bibinfo{year}{2006}).

\bibitem[{OOM()}]{OOMMF}
\bibinfo{note}{{\bf OOMMF User's Guide, Version 1.0} M.J. Donahue and D.G.
  Porter Interagency Report {\bf NISTIR 6376}, National Institute of Standards
  and Technology, Gaithersburg, MD (Sept 1999) (http://math.nist.gov/oommf/)}.

\bibitem[{\citenamefont{Cash and Karp}(1990)}]{Cash90}
\bibinfo{author}{\bibfnamefont{J.}~\bibnamefont{Cash}} \bibnamefont{and}
  \bibinfo{author}{\bibfnamefont{A.}~\bibnamefont{Karp}}, \bibinfo{journal}{ACM
  Trans. Math. Softw.} \textbf{\bibinfo{volume}{16}}, \bibinfo{pages}{201}
  (\bibinfo{year}{1990}).

\bibitem[{\citenamefont{McMichael and Stiles}(2005)}]{McMichaelJAP2005}
\bibinfo{author}{\bibfnamefont{R.~D.} \bibnamefont{McMichael}}
  \bibnamefont{and} \bibinfo{author}{\bibfnamefont{M.~D.}
  \bibnamefont{Stiles}}, \bibinfo{journal}{J. Appl. Phys.}
  \textbf{\bibinfo{volume}{97}}, \bibinfo{pages}{10J901}
  (\bibinfo{year}{2005}).

\bibitem[{\citenamefont{Bolte et~al.}(2006)\citenamefont{Bolte, Meier, and
  Bayer}}]{BayerPRB2006}
\bibinfo{author}{\bibfnamefont{M.}~\bibnamefont{Bolte}},
  \bibinfo{author}{\bibfnamefont{G.}~\bibnamefont{Meier}}, \bibnamefont{and}
  \bibinfo{author}{\bibfnamefont{C.}~\bibnamefont{Bayer}},
  \bibinfo{journal}{Phys. Rev. B} \textbf{\bibinfo{volume}{73}},
  \bibinfo{pages}{052406} (\bibinfo{year}{2006}).

\bibitem[{\citenamefont{Nibarger et~al.}(2003)\citenamefont{Nibarger, Lopusnik,
  and Silva}}]{Nibarger03}
\bibinfo{author}{\bibfnamefont{J.}~\bibnamefont{Nibarger}},
  \bibinfo{author}{\bibfnamefont{R.}~\bibnamefont{Lopusnik}}, \bibnamefont{and}
  \bibinfo{author}{\bibfnamefont{T.}~\bibnamefont{Silva}},
  \bibinfo{journal}{Appl. Phys. Lett.} \textbf{\bibinfo{volume}{82}},
  \bibinfo{pages}{2112} (\bibinfo{year}{2003}).

\bibitem[{\citenamefont{Schneider et~al.}(2005)\citenamefont{Schneider,
  Gerrits, Kos, and Silva}}]{Schneider05}
\bibinfo{author}{\bibfnamefont{M.}~\bibnamefont{Schneider}},
  \bibinfo{author}{\bibfnamefont{T.}~\bibnamefont{Gerrits}},
  \bibinfo{author}{\bibfnamefont{A.}~\bibnamefont{Kos}}, \bibnamefont{and}
  \bibinfo{author}{\bibfnamefont{T.}~\bibnamefont{Silva}},
  \bibinfo{journal}{Appl. Phys. Lett.} \textbf{\bibinfo{volume}{87}},
  \bibinfo{pages}{072509} (\bibinfo{year}{2005}).

\end{thebibliography}

\end{document}